\documentclass{article}
\usepackage{graphicx}
\usepackage{multirow}
\usepackage{amssymb}

\usepackage{authblk}

\usepackage[backend=biber,style=science,maxbibnames=20]{biblatex}

\usepackage{hyperref}

\newcommand{\data}[1]{
\section*{Data availability}
\if@anonymous Removed for anonymity \else #1 \fi}

\newcommand{\ack}[1]{
\section*{Acknowledgments}
\if@anonymous Removed for anonymity \else #1 \fi}

\newcommand{\email}[1]{\vspace*{12pt}{\fontsize{8}{10}\selectfont
       \raggedright {\bfseries E-mail:} \if@anonymous \phantom{#1} \else #1 \fi}
	  \vspace{3mm} }

\begin{document}

\title{A Thematic Analysis of A-level Physics Examiner Reports on Gravity}
\author[1]{Corey McInerney}

\affil[1]{School of Engineering and Physical Sciences, University of Lincoln, Lincoln, United Kingdom}

\maketitle

\email{cmcinerney@lincoln.ac.uk}

\begin{abstract}
Examiner reports from six major UK exam boards published between 2017 and 2025 are analysed using a mixed-methods thematic analysis. Focusing on questions relating to gravity, the objective is to understand where students commonly drop marks. Findings reveal that the source of difficulty is dependent upon topic and question style. Mathematical errors account for the majority of lost marks in calculation-type questions, while a lack of conceptual understanding commonly results in lost marks on questions relating to fields, energy and gravitational potential. Pedagogical strategies for teaching gravity must emphasise algebraic skills for orbital mechanics topics while prioritising qualitative modelling and precise definitions for field theory.
\end{abstract}

\section{Introduction}
\label{sec:Intro}

In the United Kingdom (UK), the Advanced Level (A-level) Physics course offered as part of post-16 education is a requirement for many STEM undergraduate degrees. Despite this there has been a concerning trend of declining enrolment in upper secondary physics courses seen across the UK\textsuperscript{\cite{GiBe2013}}. However, a recent curriculum assessment performed by the English government identified that enrolment onto A-level Physics courses has increased from 4\% of total A-level entries in 2009/10, to 5\% in 2023/24\textsuperscript{\cite{Curriculum2025}}. While this is positive, physics is still the least popular of the sciences at this level. 

One reason for the decline is that students often perceive physics as being difficult and unenjoyable\textsuperscript{\cite{Havard1996}}. Yet, this is a potential consequence of those students being taught by non-specialists. It was found that the majority of science teachers in England came from a biology background, therefore, those teaching physics are potentially teaching material that is less familiar to them\textsuperscript{\cite{OsSiCo2003}}. This results in inadequate explanations of scientific phenomena and unsatisfactory answers to student questions, leading to frustration and misconception\textsuperscript{\cite{AsNaAp2023}}.

The perception of difficulty is not merely a result of pedagogical delivery but is perhaps intrinsic to the subject matter itself, characterized by the presence of `threshold concepts'\textsuperscript{\cite{MeLa2006}}. Comprehension of these concepts often lead to deeper understanding of the subject as a whole\textsuperscript{\cite{WiPrWi2016}}.

One such threshold concept is gravity\textsuperscript{\cite{SmTr1988}}. Despite interacting with the Earth's gravitational field every day, research suggests that globally, students remain confused by the concept of gravity and associated ideas even through to undergraduate level\textsuperscript{\cite{BaSlDo2025, CaFa2021, WiWi2012}}. There are often discrepancies between how students describe a gravitational phenomenon and the formally accepted scientific explanation, compounded by misconceptions held by teachers and educators\textsuperscript{\cite{Syuhendri2019}}. These misconceptions can arise through a variety of factors such as: inadequate teaching material\textsuperscript{\cite{DePrSu2021}}, a difference between instruction and natural development\textsuperscript{\cite{GaBa1997}}, or even unproductive framing of the material\textsuperscript{\cite{HuHa2010}}.

Several approaches have been suggested to combat this issue, including visits to science centres\textsuperscript{\cite{Le2014}}, the use of semantics\textsuperscript{\cite{StRoMu2021}}, implementing bridging analogies\textsuperscript{\cite{AbErYi2001}} and the teaching of Einsteinian ideas instead of Newtonian gravity\textsuperscript{\cite{MacIsaac2026}}. 

In England, students are first introduced to the concept of gravity in the penultimate year of primary school. Initially introduced as the force which causes objects to fall when dropped, this notion is expanded at Key Stage 3 \& 4 level when non-contact forces are explored and gravity is explained as a field produced by all objects with mass\textsuperscript{\cite{KS1-4}}. Yet, even with the early introduction of this material, students still hold misconceptions throughout their secondary level schooling\textsuperscript{\cite{McSu2024}}. 

By reviewing examiner reports from upper secondary level physics exams across the UK, the aim of this paper is to identify what issues are holding students back from gaining marks, as directed by the exam papers' mark scheme. Are misconceptions about gravity causing students to drop marks, or is it mathematical issues where students commonly make mistakes?

The data used for this study is described in the following section. The methodologies employed are detailed in \S \ref{sec:Method} before results are presented and discussed in \S \ref{sec:R&D}.

\section{Data}
\label{sec:Data}

The dataset used consists of 137 examination questions, corresponding mark schemes and examiner reports drawn from six major UK exam boards: AQA, OCR, and Pearson (EdExcel) from England; WJEC from Wales; CCEA from Northern Ireland (NI); and SQA from Scotland.

The question papers used were selected from each county's most recent curriculum and span from 2017 to 2025 depending upon the country. Only questions relating directly to gravity and gravitational phenomenon, which have been addressed in the examiner reports have been considered in this work. As such, this dataset does not represent a complete collection of gravitational themed questions in UK post-16 education. 

Similarly, the exam paper's, mark schemes and examiner reports were all downloaded from the publicly available repository on the relevant exam board websites. Thus, if a particular document is not available on the website, that years exam was not included in this work. 

Full details of the exam paper's used can be found in Table \ref{tab:Data}.

\begin{table}[htbp]
  \centering
  \small
  \addtolength{\leftskip} {-2cm}
  \addtolength{\rightskip}{-2cm}
  \caption{Number of questions per exam paper. The `M' and `U' in the OCR papers denote two different component's (Modelling Physics and Unified Physics) which are examined separately, yet each assess the topic of gravity.}
    \begin{tabular}{|c|c|c|c|c|c|}
    \hline
    \multicolumn{1}{|c|}{\textbf{Country}} & \multicolumn{1}{c|}{\textbf{Exam Board}} & \multicolumn{1}{c|}{\textbf{Qualification}} & \multicolumn{1}{c|}{\textbf{Level}} & \multicolumn{1}{c|}{\textbf{Paper}} & \multicolumn{1}{c|}{\textbf{No. Questions}} \\
    \hline
    \multicolumn{1}{|c|}{\multirow{26}[0]{*}{England}} & \multicolumn{1}{c|}{\multirow{7}[0]{*}{AQA}} & \multicolumn{1}{c|}{\multirow{7}[0]{*}{A-Level}} & \multicolumn{1}{c|}{\multirow{7}[0]{*}{A2}} & \multicolumn{1}{c}{2017} & \multicolumn{1}{|c|}{8} \\\cline{5-6}
          &       &       &       & \multicolumn{1}{c}{2019} & \multicolumn{1}{|c|}{2} \\\cline{5-6}
          &       &       &       & \multicolumn{1}{c}{2020} & \multicolumn{1}{|c|}{5} \\\cline{5-6}
          &       &       &       & \multicolumn{1}{c}{2021} & \multicolumn{1}{|c|}{5} \\\cline{5-6}
          &       &       &       & \multicolumn{1}{c}{2022} & \multicolumn{1}{|c|}{7} \\\cline{5-6}
          &       &       &       & \multicolumn{1}{c}{2023} & \multicolumn{1}{|c|}{5} \\\cline{5-6}
          &       &       &       & \multicolumn{1}{c}{2024} & \multicolumn{1}{|c|}{3} \\\cline{2-4}\cline{5-6}
          & \multicolumn{1}{c|}{\multirow{7}[0]{*}{EdExcel}} & \multicolumn{1}{c|}{\multirow{7}[0]{*}{A-Level}} & \multicolumn{1}{c|}{\multirow{7}[0]{*}{A2}} & \multicolumn{1}{c|}{2017} & \multicolumn{1}{c|}{2} \\ \cline{5-6}
          &       &       &       & \multicolumn{1}{c}{2018} & \multicolumn{1}{|c|}{3} \\\cline{5-6}
          &       &       &       & \multicolumn{1}{c}{2020} & \multicolumn{1}{|c|}{2} \\\cline{5-6}
          &       &       &       & \multicolumn{1}{c}{2021} & \multicolumn{1}{|c|}{3} \\\cline{5-6}
          &       &       &       & \multicolumn{1}{c}{2022} & \multicolumn{1}{|c|}{1} \\\cline{5-6}
          &       &       &       & \multicolumn{1}{c}{2023} & \multicolumn{1}{|c|}{1} \\\cline{5-6}
          &       &       &       & \multicolumn{1}{c}{2024} & \multicolumn{1}{|c|}{0} \\\cline{2-4}\cline{5-6}
          & \multicolumn{1}{c|}{\multirow{12}[0]{*}{OCR}} & \multicolumn{1}{c|}{\multirow{12}[0]{*}{A-Level}} & \multicolumn{1}{c|}{\multirow{12}[0]{*}{A2}} & \multicolumn{1}{c|}{2017M} & \multicolumn{1}{c|}{2} \\ \cline{5-6}
          &       &       &       & \multicolumn{1}{c}{2018M} & \multicolumn{1}{|c|}{6} \\\cline{5-6}
          &       &       &       & \multicolumn{1}{c}{2019M} & \multicolumn{1}{|c|}{2} \\\cline{5-6}
          &       &       &       & \multicolumn{1}{c}{2020M} & \multicolumn{1}{|c|}{1} \\\cline{5-6}
          &       &       &       & \multicolumn{1}{c}{2021M} & \multicolumn{1}{|c|}{1} \\\cline{5-6}
          &       &       &       & \multicolumn{1}{c}{2022M} & \multicolumn{1}{|c|}{4} \\\cline{5-6}
          &       &       &       & \multicolumn{1}{c}{2024M} & \multicolumn{1}{|c|}{5} \\\cline{5-6}
          &       &       &       & \multicolumn{1}{c}{2018U} & \multicolumn{1}{|c|}{4} \\\cline{5-6}
          &       &       &       & \multicolumn{1}{c}{2020U} & \multicolumn{1}{|c|}{2} \\\cline{5-6}
          &       &       &       & \multicolumn{1}{c}{2021U} & \multicolumn{1}{|c|}{3} \\\cline{5-6}
          &       &       &       & \multicolumn{1}{c}{2022U} & \multicolumn{1}{|c|}{2} \\\cline{5-6}
          &       &       &       & \multicolumn{1}{c}{2023U} & \multicolumn{1}{|c|}{4} \\\cline{5-6}
          \hline
    \multicolumn{1}{|c|}{\multirow{6}[0]{*}{Northern Ireland}} & \multicolumn{1}{c|}{\multirow{6}[0]{*}{CCEA}} & \multicolumn{1}{c|}{\multirow{6}[0]{*}{A-Level}} & \multicolumn{1}{c|}{\multirow{6}[0]{*}{A2}} & \multicolumn{1}{c|}{2018} & \multicolumn{1}{c|}{4} \\
    \cline{5-6}
          &       &       &       & \multicolumn{1}{c}{2019} & \multicolumn{1}{|c|}{4} \\\cline{5-6}
          &       &       &       & \multicolumn{1}{c}{2022} & \multicolumn{1}{|c|}{3} \\\cline{5-6}
          &       &       &       & \multicolumn{1}{c}{2023} & \multicolumn{1}{|c|}{3} \\\cline{5-6}
          &       &       &       & \multicolumn{1}{c}{2024} & \multicolumn{1}{|c|}{3} \\\cline{5-6}
          &       &       &       & \multicolumn{1}{c}{2025} & \multicolumn{1}{|c|}{5} \\\cline{5-6}
    \hline
    \multicolumn{1}{|c|}{\multirow{5}[0]{*}{Wales}} & \multicolumn{1}{c|}{\multirow{5}[0]{*}{WJEC}} & \multicolumn{1}{c|}{\multirow{5}[0]{*}{A-Level}} & \multicolumn{1}{c|}{\multirow{5}[0]{*}{A2}} & \multicolumn{1}{c|}{2017} & \multicolumn{1}{c|}{5} \\ \cline{5-6}
          &       &       &       & \multicolumn{1}{c}{2019} & \multicolumn{1}{|c|}{8} \\\cline{5-6}
          &       &       &       & \multicolumn{1}{c}{2022} & \multicolumn{1}{|c|}{6} \\\cline{5-6}
          &       &       &       & \multicolumn{1}{c}{2023} & \multicolumn{1}{|c|}{3} \\\cline{5-6}
          &       &       &       & \multicolumn{1}{c}{2024} & \multicolumn{1}{|c|}{2} \\\cline{5-6}
    \hline
    \multicolumn{1}{|c|}{\multirow{5}[0]{*}{Scotland}} & \multicolumn{1}{c|}{\multirow{5}[0]{*}{SQA}} & \multicolumn{1}{c|}{\multirow{5}[0]{*}{Higher}} & \multicolumn{1}{c|}{\multirow{2}[0]{*}{Higher}} & \multicolumn{1}{c}{2024} & \multicolumn{1}{|c|}{3} \\\cline{5-6}
          &       &       &       & \multicolumn{1}{c}{2025} & \multicolumn{1}{|c|}{2} \\\cline{4-4}\cline{5-6}
          &       &       & \multicolumn{1}{c|}{\multirow{3}[0]{*}{Advanced Higher}} & \multicolumn{1}{c|}{2023} & \multicolumn{1}{c|}{2} \\\cline{5-6}
          &       &       &       & \multicolumn{1}{c}{2024} & \multicolumn{1}{|c|}{1} \\\cline{5-6}
          &       &       &       & \multicolumn{1}{c}{2025} & \multicolumn{1}{|c|}{1} \\\cline{5-6}
          \hline
          \multicolumn{5}{|c|}{\textbf{Total}}             & 138 \\
          \hline
    \end{tabular}
  \label{tab:Data}
\end{table}

\subsection{The Exam Boards \& Specifications}
\label{sec:ExamBoards}

All questions in the dataset originate from specifications aligned with national curriculum frameworks in the exam boards respective countries, which adhere to a core physics curriculum mandated by their respective regulatory bodies. Thus, the gravitational content in each board specification is governed by broadly comparable learning outcomes despite potential differences in specification structure and assessment style. The mathematical and conceptual treatment of gravity is largely invariant across exam boards. 

A critical distinction must be drawn between the qualifications offered by the English/Welsh/NI boards and the Scottish authority. The GCE A-Level (AQA\textsuperscript{\cite{AQA}}, EdExcel\textsuperscript{\cite{EdExcel}}, OCR\textsuperscript{\cite{OCR}}, WJEC\textsuperscript{\cite{WJEC}}, CCEA\textsuperscript{\cite{CCEA}}) is a two-year linear course, with gravitational fields typically taught in the second year (Year 13), which is the final year of secondary education in the UK. These specifications require a treatment of gravitational fields, gravitational potential, Kepler's Third Law and gravitational energy.

In contrast, the SQA framework is split into two tiers, examined at different stages:

\begin{itemize}
    \item[Higher\textsuperscript{\cite{SQA_H}} (H)] Generally taken in S5 (Year 12 equivalent). The gravity content here is introductory, focusing on Newton’s Law of Universal Gravitation, motion and free fall.
    
    \item[Advanced Higher\textsuperscript{\cite{SQA_AH}} (AH)] Taken in S6 (Year 13 equivalent). This specification aligns closely with the A-Level standard, introducing field theory, gravitational potential, and escape velocity calculations.
\end{itemize}

Table \ref{tab:Content} details the examinable content from each exam board.

\begin{table}[htbp]
  \centering
  \addtolength{\leftskip} {-2cm}
  \addtolength{\rightskip}{-2cm}
  \caption{Exam board content relating to gravity.}
    \begin{tabular}{l|l|l|l|l|l|l|l|}
    \cline{2-8}
          & \textbf{AQA}   & \textbf{OCR}   & \textbf{EdExcel} & \multicolumn{1}{|c|}{\textbf{CCEA}} & \multicolumn{1}{|c|}{\textbf{WJEC}} & \multicolumn{1}{|c|}{\textbf{H}} & \multicolumn{1}{|c|}{\textbf{AH}} \\
          \hline
    \multicolumn{1}{|l|}{Newton's Universal Law of Gravitation} & \checkmark     & \checkmark     & \checkmark     & \multicolumn{1}{c|}{\checkmark} & \multicolumn{1}{c|}{\checkmark} & \multicolumn{1}{c|}{\checkmark} & \multicolumn{1}{c|}{\checkmark} \\
    \hline
    \multicolumn{1}{|l|}{Gravitational field strength} & \checkmark     & \checkmark     & \checkmark     & \multicolumn{1}{c|}{\checkmark} &       &       & \multicolumn{1}{c|}{\checkmark} \\
    \hline
    \multicolumn{1}{|l|}{Gravitational force} & \checkmark     & \checkmark     & \checkmark     & \multicolumn{1}{c|}{\checkmark} & \multicolumn{1}{c|}{\checkmark} & \multicolumn{1}{c|}{\checkmark} & \multicolumn{1}{c|}{\checkmark} \\
    \hline
    \multicolumn{1}{|l|}{Gravitational potential} & \checkmark     & \checkmark     & \checkmark     &       &       &       & \multicolumn{1}{c|}{\checkmark} \\
    \hline
    \multicolumn{1}{|l|}{Orbits \& satellites} & \checkmark     & \checkmark     & \checkmark     & \multicolumn{1}{c|}{\checkmark} & \multicolumn{1}{c|}{\checkmark} & \multicolumn{1}{c|}{\checkmark} &  \\
    \hline
    \multicolumn{1}{|l|}{Kepler's Third Law} & \checkmark     & \checkmark     &       & \multicolumn{1}{c|}{\checkmark} & \multicolumn{1}{c|}{\checkmark} &       &  \\
    \hline
    \multicolumn{1}{|l|}{Escape velocity} & \checkmark     & \checkmark     &       &       &       &       & \multicolumn{1}{c|}{\checkmark} \\
    \hline
    \multicolumn{1}{|l|}{Gravitational energy} & \checkmark     & \checkmark     & \checkmark     &       &       &       & \multicolumn{1}{c|}{\checkmark} \\
    \hline
    \multicolumn{1}{|l|}{Work done} & \checkmark     & \checkmark     & \checkmark     &       &       &       &  \\
    \hline
    \end{tabular}
  \label{tab:Content}
\end{table}

Examiner reports from each board are produced by senior examiners with responsibility for standardisation and post-series review. These reports are designed to summarise candidate performance for the exam series rather than focus on individual exam scripts and thus provide a robust, board-specific but conceptually comparable source of diagnostic information. By incorporating reports from multiple awarding bodies, the present study mitigates the risk that observed misconceptions arise from idiosyncratic assessment design or local pedagogical practices tied to a single specification.

In methodological terms, exam board identity was retained as a variable in the dataset but was not used as a primary analytical dimension in this study. This decision reflects the research aim of identifying cross-specification conceptual difficulties rather than board-specific performance trends. However, the inclusion of multiple boards strengthens the generalisability of findings by ensuring that identified misconception families are not confined to a single awarding organisation’s assessment practices.

\section{Method}
\label{sec:Method}

This study employs a mixed-methods approach, combining quantitative statistical analysis of assessment materials with a qualitative thematic analysis of examiner feedback to identify recurring student mistakes and infer underlying misconceptions.

\subsection{Question Coding}
\label{sec:QuestionCoding}

Each question was manually coded by topic and type/style using a single dominant classification. The topic classifications are based on the dominant concept being examined in the question. Where overlap between concepts occurred, the coding reflects the concepts which made up the majority of the assessed concept. For example, if a question asked to calculate the distance between a star and a planet and the mark scheme makes use of Kepler's Third Law, then that question would be coded as `Kepler'. The full list of topic classifications is given in Table \ref{tab:Topic}. 

\begin{table}[htbp]
  \centering
  \addtolength{\leftskip} {-2cm}
  \addtolength{\rightskip}{-2cm}
  \caption{Exam questions by topic.}
    \begin{tabular}{|l|c|c|c|c|c|c|c|c|}
    \hline
    & \multicolumn{3}{|c|}{\textbf{England}} & \multicolumn{1}{c|}{\textbf{NI}} & \multicolumn{1}{c|}{\textbf{Wales}} & \multicolumn{2}{c|}{\textbf{Scotland}} &  \\
    \cline{2-8}
    \multicolumn{1}{|c|}{\textbf{Topic}} & \multicolumn{1}{c|}{\textbf{AQA}} & \multicolumn{1}{c|}{\textbf{OCR}} & \multicolumn{1}{c|}{\textbf{EdExcel}} & \multicolumn{1}{c|}{\textbf{CCEA}} & \multicolumn{1}{c|}{\textbf{WJEC}} & \multicolumn{1}{c|}{\textbf{H}} & \multicolumn{1}{c|}{\textbf{AH}} & \multicolumn{1}{c|}{\textbf{Total}}  \\
    \hline
    \multicolumn{1}{|l|}{Orbit} & 9     & 13    & 9     & 10    & 7     &      &      & 48 \\
    \hline
    \multicolumn{1}{|l|}{Field} & 10    & 6     &      & 2     & 5     &      &      & 23 \\
    \hline
    \multicolumn{1}{|l|}{Potential} & 6     & 3     & 3     &      & 3     &      & 2     & 17 \\
    \hline
    \multicolumn{1}{|l|}{Force} & 2     & 1     &      & 3     & 5     & 3     &      & 14 \\
    \hline
    \multicolumn{1}{|l|}{Kepler} & 6     & 4     &      & 4     & 3     &      &      & 17 \\
    \hline
    \multicolumn{1}{|l|}{Energy} &      & 7     &      &      & 1     & 1     &      & 9 \\
    \hline
    \multicolumn{1}{|l|}{Escape Velocity} & 1     &   1   &      &      &      &      & 2     & 4 \\
    \hline
    \multicolumn{1}{|l|}{Newton} &      &      &      & 1     &      &      &      & 1 \\
    \hline
    \multicolumn{1}{|l|}{Acceleration} &      & 1     &      & 2     &      &      &      & 3 \\
    \hline
    \multicolumn{1}{|l|}{Motion} & 1     &      &      &      &      &      &      & 1 \\
    \hline
    \multicolumn{1}{|c|}{\textbf{Total}}      & 35    & 36    & 12    & 22    & 24    & 4     & 4     & 137 \\
    \hline
    \end{tabular}
  \label{tab:Topic}
\end{table}

The question type classifications are based on what the questions asks the student to do or where the majority of the marks will come from. A question which explicitly states `calculate the escape velocity of object X from the surface of Mars' would be coded as `calculation'. The coded question types are given in Table \ref{tab:Type}. 

\begin{table}[htbp]
  \centering
  \addtolength{\leftskip} {-2cm}
  \addtolength{\rightskip}{-2cm}
  \caption{Exam questions by type}
    \begin{tabular}{|l|c|c|c|c|c|c|c|c|}
    \hline
    & \multicolumn{3}{|c|}{\textbf{England}} & \multicolumn{1}{c|}{\textbf{NI}} & \multicolumn{1}{c|}{\textbf{Wales}} & \multicolumn{2}{c|}{\textbf{Scotland}} &  \\
    \cline{2-8}
    \multicolumn{1}{|c|}{\textbf{Type}} & \multicolumn{1}{c|}{\textbf{AQA}} & \multicolumn{1}{c|}{\textbf{OCR}} & \multicolumn{1}{c|}{\textbf{EdExcel}} & \multicolumn{1}{c|}{\textbf{CCEA}} & \multicolumn{1}{c|}{\textbf{WJEC}} & \multicolumn{1}{c|}{\textbf{H}} & \multicolumn{1}{c|}{\textbf{AH}} & \multicolumn{1}{c|}{\textbf{Total}}  \\
    \hline
    \multicolumn{1}{|l|}{Define} & 3     & \multicolumn{1}{c|}{1} & \multicolumn{1}{c|}{} & 4     & 1     &      &      & 9 \\
    \hline
    \multicolumn{1}{|l|}{Explain} & 13    & \multicolumn{1}{c|}{9} & \multicolumn{1}{c|}{1} & 3     & 8     & 1     & 2     & 37 \\
    \hline
    \multicolumn{1}{|l|}{Calculation} & 15    & \multicolumn{1}{c|}{19} & \multicolumn{1}{c|}{11} & 14    & 12    & 3     & 2     & 76 \\
    \hline
    \multicolumn{1}{|l|}{Explain \& Calculate} &      & \multicolumn{1}{c|}{1} & \multicolumn{1}{c|}{} &      &      &      &      & 1 \\
    \hline
    \multicolumn{1}{|l|}{Derive} & 2     & \multicolumn{1}{c|}{3} & \multicolumn{1}{c|}{} & 1     & 1     &      &      & 7 \\
    \hline
    \multicolumn{1}{|l|}{Sketch} & 2     & \multicolumn{1}{c|}{3} & \multicolumn{1}{c|}{} &      & 2     &      &      & 7 \\
    \hline
    \multicolumn{1}{|c|}{\textbf{Total}}      & 35    & 36    & 12    & 22    & 24    & 4     & 4     & 137 \\
    \hline
    \end{tabular}
  \label{tab:Type}
\end{table}

\subsection{Thematic Coding}
\label{sec:ThemeCoding}

Examiner reports were analysed using inductive thematic coding following the framework outlined by Naeem et al., 2023\textsuperscript{\cite{NaOzHo2023}}. First, all report comments were read by a single researcher for consistency and a selection of recurring terms and phrases were identified. Then, descriptive codes based on these term and phrases were developed by the same researcher and assigned to relevant examiner comments. Next, the coded comments were grouped based on overarching theme and analysis carried out.

The code, groups and terms/comments/phrases upon which they are based are listed in Table \ref{tab:Key} alongside examples of when these codes would be applied in the examiner reports.

\begin{table}[htbp]
    \centering
    \addtolength{\leftskip} {-2cm}
    \addtolength{\rightskip}{-2cm}
    \caption{Coding used.}
    \begin{tabular}{|c|c|p{2in}|p{2in}|}
    \hline
    \textbf{Grouped Code}    & \textbf{Code}    & \multicolumn{1}{c|}{\textbf{Reasoning}}   & \multicolumn{1}{c|}{\textbf{Example}} \\
    \hline
    \multirow{5}{*}{Conceptual Gap}     & \multirow{2}{*}{Wrong explanation}   & Wrong physics                                                   & `... their reasoning was sometimes false.'                                                                     \\
    \cline{3-4}
                                        &                                      & Rewording/restating question                                    & `... answers simply reinterpreted the words of the question'                                                   \\
    \cline{2-4}
    & \multirow{3}{*}{Wrong definition}    & Incorrect sign                                                  & `Almost all candidates forgot that Gravitational Potential Energy is negative.'                                \\
    \cline{3-4}
                                        &                                      & Incorrect definition                                            & `The main difficulty here was remembering which was Kepler’s 1st Law.'                                         \\
    \cline{3-4}
                                        &                                      & Using the wrong term                                            & `... a few students referred to gravitational force when they were really referring to gravitational field.'   \\
    \hline
    \multirow{6}{*}{Mathematical Error} & \multirow{2}{*}{Wrong equation}      & Using the wrong equation                                        & `... many candidates failed to select the correct equation from the data booklet.'                             \\
    \cline{3-4}
                                        &                                      & Omission of/incorrect indices                                   & `... miss that the $\pi$ term was squared'                                                                         \\
    \cline{2-4}
    & \multirow{2}{*}{Algebra}             & Making mistakes in algebra: rearranging, vector addition etc.   & `A few candidates ‘dropped’ the square sign at the substitution stage...'                                      \\
    \cline{3-4}
                                        &                                      & Wrong substitution of data                                      & `... a few used the value for g rather than G.'                                                                \\
    \cline{2-4}
    & Significant figures                              & Not giving answer to requested level of significant figures     & `A significant number that were correct failed to answer using  3 significant figures...'                      \\
    \cline{2-4}
    & Units                                & Wrong/no unit conversion                                        & `The unit km was not always converted to m.'                                                                   \\
    \hline
    \multirow{5}{*}{Exam Technique}     & Verbose                              & Writing as much as possible, hoping some of it will gain a mark & `... responses were characterised by a sprawl of figures leaving the examiner to hunt for appropriate values.' \\
    \cline{2-4}
    & \multirow{2}{*}{Insufficient detail} & Not enough detail/vague answers                                 & Only a minority of students gave a complete answer.'                                                           \\
    \cline{3-4}
                                        &                                      & Missing information                                             & `Some candidates failed to mention ...'                                                                        \\
    \cline{2-4}
    & \multirow{2}{*}{Misunderstanding}    & Misreading/misinterpreting the question                         & `... lost a mark for not including the detail that...'                                                         \\
    \cline{3-4}
                                        &                                      & Not answering the question asked/not using directed method      & `... many conflated these and ended up answering a question about...'                                          \\
    \hline
    Generic   & General statement                    & No indication of candidate errors                               & `Well answered in general.' \\ 
    \hline                              
    \end{tabular}
    \label{tab:Key}
\end{table}

The overall sentiment of each examiner comment was also coded as either `positive', `mixed', `negative' or `no indication'. Phrases like `most candidates were able to' and `well answered in general' were coded as positive. Overall negative comments took the form of `poorly answered', while comments were classified as mixed if they did not indicate an overall good or bad performance by candidates, indicative in phrases like `some candidates failed'. If the examiner report made no allusion to candidate performance, it was coded as `no indication'. Such occurrences often described how the mark scheme wanted the question answered, but provided no indication of candidate performance. 

\section{Results \& Discussion}
\label{sec:R&D}

\subsection{Question Types \& Topics}
\label{QuestionTypes&Topics}

Analysis of 137 examination questions shows that gravitational physics is assessed predominantly through questions relating to orbits and gravitational fields. Orbital motion constitutes the largest proportion of questions (31\%), followed by gravitational energy and potential (24\%). Gravitational fields (19\%), Keplerian relationships (12\%) and force (11\%) make up a smaller part of the sample.

This distribution reflects a curriculum emphasis on gravity as a dynamical and spatial interaction rather than as a simple force law alone. The prominence of orbit-based questions suggests that examinations prioritise the application of gravitational theory to astrophysical contexts.

\begin{table}[htbp]
    \centering
    \addtolength{\leftskip} {-2cm}
    \addtolength{\rightskip}{-2cm}
    \caption{Exam questions by type \& topic.}
    \begin{tabular}{|c|c|c|c|c|c|c|c|c|c|c|c|}
    \hline
    \multirow{2}[0]{*}{\textbf{Type}} & \multicolumn{7}{c|}{\textbf{Topic}}                    & \multicolumn{1}{c|}{\multirow{2}[0]{*}{\textbf{Total}}} \\
                  \cline{2-8}
          & \multicolumn{1}{c|}{Orbit} & \multicolumn{1}{c|}{Energy/Potential} & \multicolumn{1}{c|}{Field} & \multicolumn{1}{c|}{Kepler} & \multicolumn{1}{c|}{Force} & \multicolumn{1}{c|}{Escape Velocity} & \multicolumn{1}{c|}{Newton} &  \\
    \hline
    \multicolumn{1}{|l|}{Calculate} & \multicolumn{1}{c|}{26} & \multicolumn{1}{c|}{19} & \multicolumn{1}{c|}{11} & \multicolumn{1}{c|}{7} & \multicolumn{1}{c|}{10} & \multicolumn{1}{c|}{3} &       & 76 \\
    \hline
    \multicolumn{1}{|l|}{Explain} & \multicolumn{1}{c|}{13} & \multicolumn{1}{c|}{10} & \multicolumn{1}{c|}{7} & \multicolumn{1}{c|}{1} & \multicolumn{1}{c|}{4} & \multicolumn{1}{c|}{1} &       & 36 \\
    \hline
    \multicolumn{1}{|l|}{Define} &       & \multicolumn{1}{c|}{2} & \multicolumn{1}{c|}{3} & \multicolumn{1}{c|}{2} &       &       & \multicolumn{1}{c|}{1} & 8 \\
    \hline
    \multicolumn{1}{|l|}{Sketch} & \multicolumn{1}{c|}{3} &       & \multicolumn{1}{c|}{4} &       &       &       &       & 7 \\
    \hline
    \multicolumn{1}{|l|}{Derive} &       & \multicolumn{1}{c|}{1} &       & \multicolumn{1}{c|}{5} &       & \multicolumn{1}{c|}{1} &       & 7 \\
    \hline
    \multicolumn{1}{|l|}{Describe} &       &       & \multicolumn{1}{c|}{1} &       & \multicolumn{1}{c|}{1} &       &       & 2 \\
    \hline
    \multicolumn{1}{|l|}{State} &       &       &       & \multicolumn{1}{c|}{1} &       &       &       & 1 \\
    \hline
    \multicolumn{1}{|c|}{\textbf{Total}} & 42    & 32    & 26    & 16    & 15    & 5     & 1     & 137 \\
    \hline
    \end{tabular}
  \label{tab:TypeTopic}
\end{table}

Calculation-type questions were the most prevalent across all exam boards (55\%), followed by explanation of a statement, phenomenon or data (26\%). Table \ref{tab:TypeTopic} is a cross-tabulation of topic and question type which shows this in more detail. 



Despite this work's focus on examination errors, when examiner reports are assessed based on their overall sentiment (Table \ref{tab:TopicSentiment} provides a breakdown of question topic against report sentiment), it is found that they are mostly positive (60\%). This points towards the gravity topic being understood well enough by the majority of candidates to achieve most of the points in the mark schemes. 

While this is a favourable impression, approximately one third of comments were overall negative, signalling that a significant number of students are failing to achieve the necessary level of rigour or explain their understanding in suitable ways. 

\begin{table}[htbp]
  \centering
  \caption{Count of topic by sentiment.}
    \begin{tabular}{|c|l|c|}
    \hline
    \multicolumn{1}{|c|}{\textbf{Topic}}              & \textbf{Sentiment}     & \textbf{Count} \\
    \hline
    \multirow{4}[0]{*}{Energy/Potential} & \multicolumn{1}{l|}{Negative} & 13 \\
          & \multicolumn{1}{l|}{Positive} & 9 \\
          & \multicolumn{1}{l|}{Mixed} & 7 \\
          & \multicolumn{1}{l|}{No indication} & 3 \\
          \hline
    \multicolumn{2}{|c|}{Energy/Potential Total} & 32 \\
    \hline
    \multirow{4}[0]{*}{Orbit} & \multicolumn{1}{l|}{Positive} & 21 \\
          & \multicolumn{1}{l|}{Negative} & 10 \\
          & \multicolumn{1}{l|}{Mixed} & 9 \\
          & \multicolumn{1}{l|}{No indication} & 2 \\
          \hline
    \multicolumn{2}{|c|}{Orbit Total} & 42 \\
    \hline
    \multirow{3}[0]{*}{Kepler} & \multicolumn{1}{l|}{Positive} & 10 \\
          & \multicolumn{1}{l|}{Mixed} & 3 \\
          & \multicolumn{1}{l|}{Negative} & 3 \\
          \hline
    \multicolumn{2}{|c|}{Kepler Total} & 16 \\
    \hline
    \multirow{4}[0]{*}{Field} & \multicolumn{1}{l|}{Negative} & 13 \\
          & \multicolumn{1}{l|}{Positive} & 8 \\
          & \multicolumn{1}{l|}{Mixed} & 4 \\
          & \multicolumn{1}{l|}{No indication} & 1 \\
          \hline
    \multicolumn{2}{|c|}{Field Total} & 26 \\
    \hline
    \multirow{3}[0]{*}{Force} & \multicolumn{1}{l|}{Positive} & 9 \\
          & \multicolumn{1}{l|}{Mixed} & 3 \\
          & \multicolumn{1}{l|}{Negative} & 3 \\
          \hline
    \multicolumn{2}{|c|}{Force Total} & 15 \\
    \hline
    \multirow{3}[0]{*}{Escape Velocity} & \multicolumn{1}{l|}{Positive} & 2 \\
          & \multicolumn{1}{l|}{Negative} & 2 \\
          & \multicolumn{1}{l|}{Mixed} & 1 \\
          \hline
    \multicolumn{2}{|c|}{Escape Velocity Total} & 5 \\
    \hline
    \multicolumn{1}{|c|}{Newton} & \multicolumn{1}{l|}{Positive} & 1 \\
    \hline
    \multicolumn{2}{|c|}{Newton Total} & 1 \\
    \hline
    \end{tabular}
  \label{tab:TopicSentiment}
\end{table}

\subsubsection{Energy \& Potential}

Topics such as energy and potential have a high proportion of `negative' comments (41\%). Looking at Table \ref{tab:TypeTopic}, we see that calculation-type questions are the most common for these topics, however, mathematical errors are not where the majority of marks are lost for these topics. Table \ref{tab:TypeGroup} shows the error type against question topic and in fact, it is conceptual errors which cause the most problems. This is evidenced in comments such as `question 3(b) was a question designed to test candidates’ knowledge that gravitational potential energy is always negative ... the majority of candidates gave their answer as $(3.5 + 7) \times 10^{11}$ J rather than $(3.5 – 7) \times 10^{11}$ J.' 

Mistaking the sign in the equation for potential has been identified as leading to conceptual errors and incorrect thinking even at undergraduate level\textsuperscript{\cite{Lindsey2014}}. Teaching strategies for these topics must prioritize qualitative understanding as comments such as `some candidates were able to identify that the gravitational potential energy of Dimorphos would be less. However, only a few could give an appropriate explanation for this change' indicate an inability to go beyond textbook material. Problems such as this may be alleviated with the careful use of metaphor\textsuperscript{\cite{WiAlMi2016}}. 

\begin{table}[htbp]
  \centering
    \small
    \addtolength{\leftskip} {-2cm}
    \addtolength{\rightskip}{-2cm}
    \caption{Question type against error type.}
    \begin{tabular}{|l|c|c|c|c|c|c|c|c|c|c|c|}
    \hline
    \multicolumn{1}{|c|}{\multirow{2}[0]{*}{\textbf{Grouped Code}}} & \multicolumn{7}{c|}{\textbf{Topic}} \\
    \cline{2-8}
    & \multicolumn{1}{l|}{Orbit} & \multicolumn{1}{l|}{Energy/Potential} & \multicolumn{1}{l|}{Field} & \multicolumn{1}{l|}{Kepler} & \multicolumn{1}{l|}{Force} & \multicolumn{1}{l|}{Escape Velocity} & \multicolumn{1}{l|}{Newton} \\
    \hline
    Mathematical Error & 27    & 9     & 8     & 6     & 8     & 2     &  \\
    \hline
    Conceptual Gap & 13    & 15    & 12    & 3     & 3     & 2     & 1 \\
    \hline
    Generic & 6     & 7     & 4     & 4     & 2     & 1     &  \\
    \hline
    Exam Technique & 7     & 8     & 8     & 5     & 4     &       & 1 \\
    \hline
    \multicolumn{1}{|c|}{\textbf{Total}} & 53    & 39    & 32    & 18    & 17    & 5     & 2 \\
    \hline
    \end{tabular}
    \label{tab:TypeGroup}
\end{table}

\subsubsection{Orbits, Kepler \& Fields}

Orbits are the most commonly assessed topic in this study and have a high percentage of `positive' comments with 50\%. 26 out of the 42 orbit based questions were calculation-based, leading to 51\% of errors being mathematical, double the amount of conceptual errors. This distribution points towards an emphasis on the calculation of parameters such as mass and orbital radius. When orbit questions are not calculation based, they mostly pertain to explaining the motion of orbiting bodies. This is where conceptual understanding is tested with students either confusing definitions or not using the correct physics to explain orbital phenomenon.

Focusing on the qualitative aspects of motion before introducing quantitative calculations has been shown to improve student understanding and performance\textsuperscript{\cite{Prescott2004}}, an approach which could prove useful in the teaching of orbital mechanics. 

Questions relating to Kepler's Laws have the highest percentage of `positive' comments in this study (63\%), consistent with the literature which has illustrated the ease with which students adopt the ideas presented by these laws\textsuperscript{\cite{YuSaDe2010, SaKu2014}}. The spread of errors for this topic is fairly evenly spread across all grouped errors, with mathematical errors being slightly more common. This is expected considering this topic is tested using 5 of the 8 question types identified, perhaps due to the interesting mathematical relationships presented, alongside the opportunity to test knowledge of definitions and explanations of the relationship between orbital parameters. 

50\% of questions relating to gravitational fields were coded as `negative', the highest of any of the topics identified. 38\% of field question comments pointed to conceptual errors as the reason for dropped marks. Examiner comments pointed out that students `referred to gravitational force when they were really referring to gravitational field', or made errors such as drawing `arrows in totally the wrong direction, sometimes placing them at right angles to the field lines and sometimes having them point away from the masses' in sketch-style questions. 

The difficulty students experience in being able to distinguish between concepts and envision abstract notions can be alleviated with the correct use of computer visualisation software such as GeoGebra and VPython\textsuperscript{\cite{MaMiBa2016}}. This type of interdisciplinary learning enables students to get hands-on with phenomenon not normally demonstrable in the classroom and allows them to interact with models and mathematics, noting and observing cause and effect. 

\subsection{Thematic Analysis}
\label{ThemeaticAnalysis}

\begin{table}[htbp]
    \centering
    \addtolength{\leftskip} {-2cm}
    \addtolength{\rightskip}{-2cm}
    \caption{Count of grouped codes by overall sentiment.}
    \begin{tabular}{|c|c|c|}
    \hline
    \textbf{Grouped Codes} & \textbf{Sentiment} & \multicolumn{1}{c|}{\textbf{Count of Grouped Codes}} \\
    \hline
    \multicolumn{1}{|c}{\multirow{3}[0]{*}{Conceptual Gap}} & \multicolumn{1}{|c|}{Negative} & 27 \\
    & \multicolumn{1}{c|}{Mixed} & 10 \\
    & \multicolumn{1}{c|}{Positive} & 9 \\
    \hline
    \multicolumn{1}{|c}{Conceptual Gap Total} &       & 46 \\
    \hline
    \multicolumn{1}{|c}{\multirow{3}[0]{*}{Exam Technique}} & \multicolumn{1}{|c|}{Negative} & 12 \\
    & \multicolumn{1}{c|}{Mixed} & 8 \\
    & \multicolumn{1}{c|}{Positive} & 12 \\
    \hline
    \multicolumn{1}{|c}{Exam Technique Total} &       & 32 \\
    \hline
    \multirow{4}[0]{*}{Mathematical Error} & \multicolumn{1}{c|}{Negative} & 17 \\
          & \multicolumn{1}{c|}{Mixed} & 13 \\
          & \multicolumn{1}{c|}{Positive} & 25 \\
          & \multicolumn{1}{c|}{No Indication} & 1 \\
    \hline
    \multicolumn{1}{|c}{Mathematical Error Total} &       & 56 \\
    \hline
    \multirow{3}[0]{*}{Generic} & \multicolumn{1}{c|}{Negative} & 1 \\
          & \multicolumn{1}{c|}{Positive} & 19 \\
          & \multicolumn{1}{c|}{No Indication} & 5 \\
    \hline
    \multicolumn{1}{|c}{Generic Total} &       & 25 \\
    \hline
    \multicolumn{2}{|c|}{\textbf{Total}} & 159 \\
    \hline
    \end{tabular}
    \label{tab:GroupsSentiment}
\end{table}

In total, 158 phrases in the examiner reports considered in this work were coded based on the rubric detailed in \S \ref{sec:ThemeCoding}. The count of grouped codes against examiner report sentiment is detailed in Table \ref{tab:GroupsSentiment}.

\subsubsection{Types of Errors}
\label{TypesOfErrors}

\begin{figure}
    \centering
    \includegraphics[width=\linewidth]{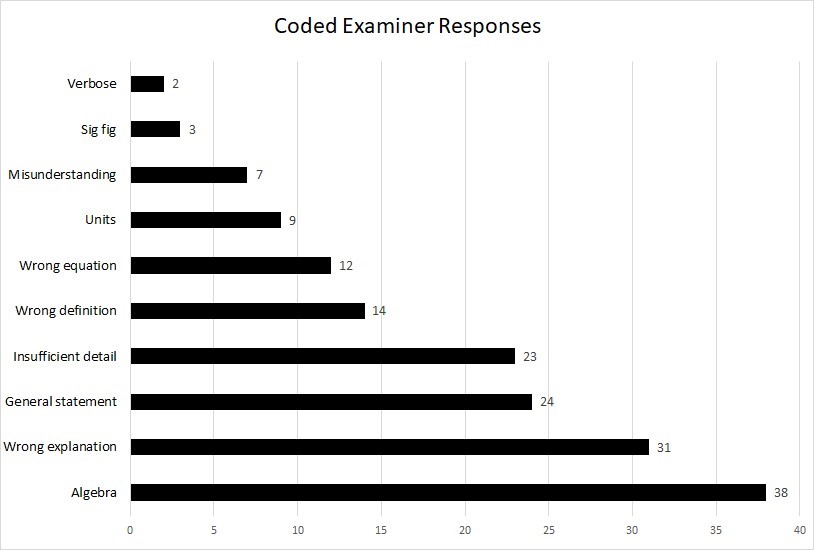}
    \caption{Count of coded statements.}
    \label{fig:Code}
\end{figure}

Over a third of the examiner reports pointed towards mathematical error as the cause of students losing marks. This is not surprising considering that the majority of question-types considered were calculation based. Figure \ref{fig:Code} shows the breakdown of coded statements outside of the groups. It is clear that algebraic issues are responsible for the majority of these mathematical errors. A common algebraic error identified by examiners is in the rearranging of equations, where students `failed to rearrange the equation' or made mistakes with indices such as `forgetting to square the radius'.


Errors such as these could be related to exam stress (particularly in the cases of `forgetting' indices) or not checking working before the end of the exam. Unfortunately, this is impossible to verify with the data available in this work. On the other hand, this result may point towards a wider issue in problem solving in physics using mathematics, evidenced by examiner comments such as `students could not cope with the quite complicated proportional relationship' and `could not correctly complete the mathematics required'. This is affirmed in the literature which indicates that without a concentrated effort in the classroom to promote the interplay between physics and mathematics, students often struggle to connect a physical and mathematical interpretation of a phenomenon\textsuperscript{\cite{NiAnGr2013, Pospiech2019}}.

Methods such as mixing syntactic and semantic reasoning have been suggested as a way to bridge this gap\textsuperscript{\cite{PaRa2024}}, and for gravity specifically, co-variational reasoning has successfully been used to integrate mathematical and physical ideas\textsuperscript{\cite{BaPaZh2020, PaGe2021}}.

Yet despite the majority of lost marks being from mathematical errors, it is conceptual errors which appear the most in `negative' reports (59\%). This indicates that when a student lacks fundamental understanding (e.g., misunderstanding field lines or potential), they rarely recover marks elsewhere in the question.

In contrast, mathematical error comments were actually associated with a positive sentiment more often (45\%) than a negative one (30\%). This suggests that examiners frequently award partial credit even if the arithmetic is wrong (`making calculation errors did give an answer to 3 significant figures, and thereby gained a mark'). On the other hand, this may be a by-product of the coding used to gauge overall examiner sentiment with a question. If the report indicated that a particular question was `very well answered', this would be coded as a positive sentiment. However, that same comment goes on to say `although a small minority insisted on trying to use the force equation and not being able to proceed'. This comment is then coded as a `wrong equation' mathematics error as this is how some students were losing marks. As such, the sentiment coding should be used as an indication of how the majority of students answered a question, whereas this study is concerned with the areas where students are failing to gain credit. 

\begin{table}[htbp]
    \centering
    \addtolength{\leftskip} {-2cm}
    \addtolength{\rightskip}{-2cm}
    \caption{Question style against error type.}
    \begin{tabular}{|l|l|l|l|l|l|}
    \hline
\multicolumn{1}{|c|}{\multirow{2}{*}{\textbf{Style}}} & \multicolumn{5}{|c|}{\textbf{Grouped Codes}}                                      \\
\cline{2-6}
\multicolumn{1}{|c|}{}                       & Mathematical Error & Conceptual Gap & Generic & Exam Technique & Total \\
\hline
Calculate                                  & 53                 & 14             & 12      & 7              & 86    \\
\hline
Explain                                    & 1                  & 18             & 6       & 15             & 40    \\
\hline
Define                                     &                    & 6              & 1       & 6              & 13    \\
\hline
Sketch                                     &                    & 5              & 2       &                & 7     \\
\hline
Derive                                     & 2                  & 3              & 2       & 2              & 9     \\
\hline
Describe                                   &                    &                &         & 2              & 2     \\
\hline
State                                      &                    &                & 1       &                & 1     \\
\hline
    \end{tabular}
    \label{tab:StyleGroup}
\end{table}

By cross-referencing question style with the error group (Table \ref{tab:StyleGroup}) we see that on explain-type questions, students often fall short through conceptual (45\%) or exam technique (37\%) errors. This is indicative of students either not understanding the physics involved, or, failing to answer the question sufficiently. A similar pattern emerges for define-type questions (50\% and 42\% respectively for conceptual and exam errors). Although this style of question makes up a small portion of our sample, it is clear that define-type questions are good at exposing conceptual errors with students `not able to quote [Kepler's Third Law] correctly' or making `errors such as choosing the wrong direction'. Likewise for exam technique errors where `most of the answers seen lacked detail'.

\section{Conclusion}
\label{sec:Conclusion}

The objective of this work is to examine common areas where secondary students in the UK lose marks on exam questions relating to gravity. Presented above is analysis of a collection of 137 exam questions relating to gravity gathered from past paper examinations from six major exam boards in the UK. 

The analysis suggests that students struggle most with the mathematical aspects of the exam (equation manipulation, unit conversions etc.). This is prevalent in
questions on Kepler's Third Law and orbits in general. 

Conceptual understanding is key in questions pertaining to gravitational potential, energy and fields, where a robust understanding of definitions and qualitative descriptions of phenomenon is key to achieving the marks outlined in the mark scheme. 

Provided here is a taxonomy of difficult concepts and procedural traps that affect student outcomes. Educators need to be aware of these and apply methods in the classroom which can combat them.

\ack{The author thanks the attendees of the ISTEP 2025 conference, conversations with whom inspired this work. Thanks also go to Phil Sutton and Helen Christodoulidi for their feedback on this manuscript. }



\data{The data that support the findings of this study are openly available in figshare at https://doi.org/10.6084/m9.figshare.31376227.} 


\section*{References}

\printbibliography[heading=none]

\end{document}